\documentclass{aa}
\usepackage{graphicx}
\begin{document}

\title{The WEBT campaign to observe \object{AO 0235+16} in the 2003--2004
observing season\thanks{For questions regarding the availability
of the data presented in this paper,
please contact the WEBT President Massimo Villata ({\tt villata@to.astro.it})}}

\subtitle{Results from radio-to-optical monitoring and XMM-Newton observations}

\author{C.~M.~Raiteri \inst{1}
\and M.~Villata \inst{1}
\and M.~A.~Ibrahimov \inst{2}
\and V.~M.~Larionov \inst{3,4}
\and M.~Kadler \inst{5}
\and H~.D.~Aller \inst{6}
\and M.~F.~Aller \inst{6}
\and Y.~Y.~Kovalev \inst{7,8}\thanks{Jansky Fellow}
\and L.~Lanteri \inst{1}
\and K.~Nilsson \inst{9}
\and I.~E.~Papadakis \inst{10,11}
\and T.~Pursimo \inst{12}
\and G.~E.~Romero \inst{13}
\and H.~Ter\"asranta \inst{14}
\and M.~Tornikoski \inst{14}
\and A.~A.~Arkharov \inst{15}
\and D.~Barnaby \inst{16}
\and A.~Berdyugin \inst{9}
\and M.~B\"ottcher \inst{17}
\and K.~Byckling \inst{18}
\and M.~T.~Carini \inst{16}
\and D.~Carosati \inst{19}
\and S.~A.~Cellone \inst{20}
\and S.~Ciprini \inst{9}
\and J.~A.~Combi \inst{13,21}
\and S.~Crapanzano \inst{1}
\and R.~Crowe \inst{22}
\and A.~Di Paola \inst{23}
\and M.~Dolci \inst{24}
\and L.~Fuhrmann \inst{1,5,25}
\and M.~Gu \inst{26}
\and V.~A.~Hagen-Thorn \inst{3,4}
\and P.~Hakala \inst{18}
\and V.~Impellizzeri \inst{5}
\and S.~Jorstad \inst{27}
\and J.~Kerp \inst{5}
\and G.~N.~Kimeridze \inst{28}
\and Yu.~A.~Kovalev \inst{8}
\and A.~Kraus \inst{5}
\and T.~P.~Krichbaum \inst{5}
\and O.~M.~Kurtanidze \inst{28}
\and A.~L\"ahteenm\"aki\inst{14}
\and E.~Lindfors \inst{9}
\and M.~G.~Mingaliev\inst{29}
\and R.~Nesci \inst{30}
\and M.~G.~Nikolashvili \inst{28}
\and J.~Ohlert \inst{31}
\and M.~Orio \inst{1}
\and L.~Ostorero \inst{32}
\and M.~Pasanen \inst{9}
\and A.~Pati \inst{33}
\and C.~Poteet \inst{16}
\and E.~Ros \inst{5}
\and J.~A.~Ros \inst{34}
\and P.~Shastri \inst{33}
\and L.~A.~Sigua \inst{28}
\and A.~Sillanp\"a\"a \inst{9}
\and N.~Smith \inst{35}
\and L.~O.~Takalo \inst{9}
\and G.~Tosti \inst{25}
\and A.~Vasileva \inst{3}
\and S.~J.~Wagner \inst{32}
\and R.~Walters \inst{16}
\and J.~R.~Webb \inst{36}
\and W.~Wills \inst{16}
\and A.~Witzel \inst{5}
\and E.~Xilouris \inst{37}
}

\offprints{C.\ M.\ Raiteri, \email{raiteri@to.astro.it}} 

\institute{INAF, Osservatorio Astronomico di Torino,
     10025 Pino Torinese (TO), Italy
\and Ulugh Beg Astronomical Inst., Academy of
     Sciences of Uzbekistan, Tashkent 700052, Uzbekistan
\and Astronomical Inst., St.-Petersburg
     State Univ., 198504 St.-Petersburg, Russia
\and Isaac Newton Inst.\ of Chile, St.-Petersburg Branch, Russia
\and Max-Planck-Institut f\"ur Radioastronomie, 53121 Bonn, Germany
\and Dept.\ of Astronomy, Univ.\ of Michigan, Ann Arbor, MI
     48109, USA 
\and National Radio Astronomy Observatory, Green Bank, WV 24944, USA
\and Astro Space Center of Lebedev Physical Inst., Moscow, 117997, Russia
\and Tuorla Astronomical Observatory, Univ.\ of Turku,
     21500 Piikki\"{o}, Finland
\and IESL, FORTH, 711 10 Heraklion, Crete, Greece
\and Physics Dept., Univ.\ of Crete, 710 03 Heraklion, Crete, Greece
\and Nordic Optical Telescope, 38700 Santa Cruz de La Palma, Spain
\and Inst.\ Argentino de Radioastronom\'{\i}a, Buenos Aires, Argentina
\and Mets\"ahovi Radio Observatory, Helsinki Univ.\ of Technology,
     02540 Kylm\"al\"a, Finland
\and Pulkovo Observatory, St.Petersburg, Russia
\and Dept.\ of Physics \& Astronomy, Western Kentucky Univ.,
     Bowling Green, KY 42104, USA
\and Dept.\ of Physics and Astronomy, Ohio Univ.,
     Athens, OH 45701-2979, USA
\and Observatory, 00014 Univ.\ of Helsinki, Finland
\and Armenzano Astronomical Observatory, Assisi, Italy
\and Facultad de Ciencias Astron\'omicas y Geof\'{\i}sicas,
     UNLP, Buenos Aires, Argentina
\and Dept.\ de F\'{\i}sica, Escuela Polit\'ecnica Superior, Univ.\ de Ja\'en,
     23071 Ja\'en, Spain
\and Dept.\ of Physics \& Astronomy, Univ.\ of Hawaii, Hilo, Hawaii 96720-4091, USA
\and INAF, Osservatorio Astronomico di Roma, 00040 Monte Porzio Catone (RM), Italy
\and INAF, Osservatorio Astronomico di Collurania Teramo, 64100 Teramo, Italy
\and Osservatorio Astronomico, Univ.\ di Perugia, 06126 Perugia, Italy
\and Korea Astronomy Observatory, Taejeon 305-348,  Korea
\and Inst.\ for Astrophysical Research, Boston Univ., Boston, MA 02215, USA
\and Abastumani Observatory, 383762 Abastumani, Georgia
\and Special Astrophysical Observatory, Karachaevo-Cherkessia, 369167, Russia
\and Dipartimento di Fisica, Univ.\ di Roma ``La Sapienza'', 00185 Roma, Italy
\and Michael Adrian Observatory, 65468 Trebur, Germany
\and Landessternwarte Heidelberg, K\"onigstuhl, 69117 Heidelberg, Germany
\and Indian Inst.\ of Astrophysics, Bangalore 560 034, India
\and Agrupaci\'o Astron\`omica de Sabadell, 08200 Sabadell, Spain
\and Dept.\ of Applied Physics \& Instrumentation, Cork Inst.\ of Technology,
     Cork, Ireland
\and SARA Observatory, Florida International Univ., Miami, FL 33199, USA
\and Inst.\ of Astronomy and Astrophysics, National Observatory of Athens, 
     11810 Athens, Greece
}

\date{Received; Accepted;}

\titlerunning{The WEBT campaign on \object{AO 0235+16} in the 2003--2004 observing season}

\authorrunning{C.\ M.\ Raiteri et al.}

\abstract{
}

\maketitle

\twocolumn[{\bf Abstract.} A multiwavelength campaign to observe the BL Lac object
\object{AO 0235+16} ($z=0.94$) was set up
by the Whole Earth Blazar Telescope (WEBT) collaboration during the observing seasons 2003--2004
and 2004--2005, involving radio, near-IR and optical photometric monitoring,
VLBA monitoring, optical spectral monitoring,
and three pointings by the
XMM-Newton satellite. Here we report on the results of the first season,
which involved the participation of 24 optical and near-IR telescopes and 4 radio telescopes, as well as
the first XMM-Newton pointing, which occurred on January 18--19, 2004.
Unpublished data from previous epochs were also collected 
(from 5 optical-NIR and 3 radio telescopes),
in order to fill the gap between the end
of the period presented in Raiteri et al.\ (\cite{rai01}) and the start of the WEBT campaign.
The contribution of the southern AGN, 2 arcsec distant from the source, is taken into account.
It is found to especially affect the blue part of the optical spectrum when the source is faint.
In the optical and near-IR the source has been very active in the last 3 years,
although it has been rather faint most of the time,
with noticeable variations of more than a magnitude over a few days.
In contrast, in the radio bands it appears to have been ``quiescent" since early 2000.
The major radio (and optical) outburst predicted to peak around February--March 2004 (with
a six month uncertainty) has not occurred yet.
When comparing our results with the historical light curves,
two different behaviours seem to characterize the optical outbursts:
only the major events present a radio counterpart.
The X-ray spectra obtained by the three EPIC detectors are well fitted by a power law
with extra-absorption at $z=0.524$;
the energy index in the 0.2--10 keV range is well constrained: $\alpha=0.645 \pm 0.028$
and the 1 keV flux density is $0.311 \pm  0.008 \, \rm \mu Jy$.
The analysis of the X-ray light curves reveals that no
significant variations occurred during the pointing. In contrast, simultaneous
dense radio monitoring
with the 100 m telescope at Effelsberg shows a $\sim 2$--3\% flux decrease in 6--7 hours,
which, if intrinsic, would imply a brightness temperature well above the Compton limit and
hence a lower limit to the Doppler factor $\delta \ga 46$.
We construct the broad-band spectral energy distribution of January 18--19, 2004 with
simultaneous radio data from Effelsberg, optical data
from the Nordic Optical Telescope (NOT), optical--UV data from
the Optical Monitor onboard XMM-Newton, and X-ray data by the EPIC instruments. 
Particular care is taken to correct data for extinction due to both
the Milky Way and the $z=0.524$ absorber. The resulting SED suggests the existence of
a bump in the UV spectral region.

\bigskip]

\section{Introduction}

BL Lac objects belong to the blazar class of active galactic nuclei (AGNs). They show
variability at all wavelengths, from the radio to the $\gamma$-ray band, on a variety of time scales.
In general, slow oscillations of the flux mean level on time scales of weeks to years have been observed,
based on which fast flares lasting as little as a few hours are superposed.
The commonly accepted scenario for blazar emission assumes a black hole surrounded by an
accretion disc, feeding a plasma jet which is oriented close
to the line of sight. This geometrical configuration can explain
the apparent superluminal motion of relativistically moving plasma down the jet
which is often derived from the analysis of radio maps.
Moreover, relativistic Doppler effects can, at least in part, explain the extremely high
brightness temperature which otherwise would result in the so-called Compton catastrophe.

The best way to study the source flux variability and to understand
the corresponding physical processes is the simultaneous
observation across the whole accessible electromagnetic spectrum.
Indeed, it is commonly accepted that the observed lower-energy flux
(from the radio band to the UV--X-ray one)
is due to synchrotron emission from the plasma jet, while the higher-energy one
(up to GeV--TeV energies) is thought to be
produced by inverse-Compton scattering of
soft photons off the same relativistic electrons producing the synchrotron radiation.
It goes without saying that the high-energy flux should thus be correlated with the low-energy one,
but the relative variability amplitude and the amount of time lag expected depend on the
particular model adopted.

Analysis of the long-term radio and optical light curves of
AO 0235+16 led to the conclusion that the main
radio outbursts have optical counterparts, and that they repeat every $5.7 \pm 0.5$ years
(Raiteri et al.\ \cite{rai01}).
This was interpreted in terms of geometrical effects due to the possible presence
of a binary black hole system (Romero et al.\ \cite{rom03}; Ostorero et al.\ \cite{ost04}).
The next outburst was predicted to peak around February--March 2004,
and a multiwavelength radio-NIR-optical
campaign has been organized by the Whole Earth Blazar Telescope
(WEBT; {\tt http://www.to.astro.it/blazars/webt/}; e.g.\ Villata et al.\ \cite{vil00}, \cite{vil02},
\cite{vil04a}, \cite{vil04b}) collaboration to follow the source behaviour.
The WEBT campaign began in
summer 2003, and was planned to continue until spring 2005.
During this long-term campaign three pointings by the XMM-Newton satellite were granted in order to
compare the behaviour of the low-energy emission with the high-energy one.
The XMM-Newton observations were scheduled on January 18--19 and August 2, 2004, and January 28, 2005.
We know from previous observations by
X-ray satellites (Einstein, Worral \& Wilkes \cite{wor90};
EXOSAT, Ghosh \& Soundararajaperumal \cite{gho95}; ROSAT, Madejski et al.\ \cite{mad96},
Comastri et al.\ \cite{com97};
ASCA, Madejski et al.\ \cite{mad96}, Junkkarinen et al.\ \cite{jun04}; RXTE, Webb et al.\ \cite{web00};
BeppoSAX, Padovani et al.\ \cite{pad04}) that
both the X-ray flux and spectral index can be very variable.

The variability of AO 0235+16 on short time scales is also of special interest.
At radio wavelengths several flat-spectrum radio quasars and BL Lacs are known to
exhibit rapid variability (Witzel et al.\ \cite{wit86}; Heeschen et al.\ \cite{hee87}; 
Kedziora-Chudczer et al.\ \cite{ked01}; Lovell et al.\ \cite{lov03};
Kraus et al.\ \cite{kra03}) on time scales of days to hours
(intraday variability, IDV). The cause of the variations seen in these sources is
currently controversial, with claims being made for either a source-intrinsic
(e.g.\ shocks in the emitting jet) or extrinsic origin (interstellar scintillation or
gravitational microlensing).
Interstellar scintillation (ISS) may play an important role in the cm-radio regime
(Jauncey \& Macquart \cite{jau01}; Rickett et al.\ \cite{ric01}; 
Dennet-Thorpe \& de Bruyn \cite{den02}), otherwise intrinsic variations
may require extremely high Doppler boosting ($\delta \sim 100$) or special source geometries in
order to prevent the inverse-Compton catastrophe (see e.g.\ Wagner \& Witzel
\cite{wag95}; Wagner et al.\ \cite{wag96}).  Since ISS cannot explain correlated IDV over a
wide range of the electromagnetic spectrum, the detection of broad-band correlations would
directly rule out ISS as sole explanation for radio IDV and would favour a source intrinsic 
contribution to the IDV pattern.

AO 0235+16 has shown IDV over a wide range of the electromagnetic spectrum, 
from the radio to the optical band (Takalo et al.\ \cite{tak92};
Heidt \& Wagner \cite{hei96}; Noble \& Miller \cite{nob96}; Romero et al.\ \cite{rom97};
Kraus et al.\ \cite{kra99}; Romero et al.\ \cite{rom00}; Raiteri et al.\ \cite{rai01}).
Moreover, hints for the existence of very high Doppler factors come from a variety of other
observational studies 
(Fujisawa et al.\ \cite{fuj99}; Frey et al.\ \cite{fre00}; Jorstad et al.\ \cite{jor01}),
thus making this blazar an ideal candidate for the detection of correlated broad-band IDV.
XMM-Newton has already revealed fast X-ray variability of AO 0235+16
(Kadler et al.\ \cite{kad05}).
Consequently, one of the aims of the WEBT campaign was to search for
radio-NIR-optical-X-ray correlated fast variations during the XMM-Newton pointings,
with the support of dense radio monitoring by the 100 m radio telescope at Effelsberg.

Side by side with the WEBT campaign, spectroscopic monitoring of the source in the optical band
with both the Telescopio Nazionale Galileo (TNG) and the ESO's New Technology Telescope (NTT)
has been carried out (Raiteri et al.\ \cite{rai05}).
15 VLBA epochs have been awarded to monitor the source structure variability
during the WEBT campaign (Wiik et al.\ \cite{wii05}).

In this paper we present results obtained during the first observing season of the campaign,
from July 1, 2003 to April 2004, including the first XMM-Newton pointing of January 18--19, 2004.
We also collected data taken before the start of the campaign,
back to the last data published by Raiteri et al.\ (\cite{rai01}), in order to reconstruct
the source behaviour in between.
Other papers will follow, containing new results and more detailed analyses.

The long-term optical, NIR and radio monitoring data obtained by the WEBT during the
2003--2004 observing season, along with the data filling the gap between this WEBT campaign
and the data published by Raiteri et al.\ (\cite{rai01}), are presented in Sect.\ 2.
The multiwavelength observations performed during the XMM-Newton pointing of January 18--19, 2004
are analysed in Sect.\ 3.
In Sect.\ 4 we discuss the problem of extinction and construct the broad-band spectral 
energy distribution of AO 0235+16
during the XMM-Newton pointing. Section 5 contains some results of the cross-correlation analysis.
Conclusions are drawn in Sect.\ 6.

\section{Long-term observations by the WEBT}

The WEBT campaign was started on July 1, 2003; 
the last data presented here were taken in April 2004.
However, for the present work we collected data from previous epochs in order to fill the gap
between the start of the WEBT campaign and the last data published in
Raiteri et al.\ ({\cite{rai01}), and to enrich the historical light curves.

\subsection{Optical and near-IR data}

Tables \ref{obs_opt} and \ref{obs_ir} show the lists of the optical and near-IR telescopes
participating in the WEBT campaign (and prior data collection) ordered by
longitude\footnote{One of the main advantages of WEBT is to
have observers spread out at different longitudes. Hence, the observing task
can move from east to west allowing us in principle to obtain continuous monitoring
over 24 hours.}.
In the tables Col.\ 1 contains the name of the observatory and its location;
Col.\ 2 gives the telescope diameter in cm; Cols.\ 3--7 report the number of observations
done in the $UBVRI$ and $JHK$ bands;
Col.\ 8 shows the total number of observations performed by each telescope in those bands.
Data in brackets indicate the number of points remaining after the ``cleaning" procedure described
below and which are shown in the light curves.
As one can see in the last row, the total number of observations collected is 2509 in the optical and
385 in the near-IR.

\begin{table*}
\centering
\caption{List of participating optical observatories by longitude and number
of observations performed in the various bands.}
\begin{tabular}{l c |c c c c c| c}
\hline
Observatory            &$d$ (cm)&$N_U$  &$N_B$    &$N_V$    &$N_R$    &$N_I$    &$N_{\rm tot}$\\
\hline
IAO Hanle, India          &200 &8 (6)  &7 (6)    &12 (10)  &8 (8)    &14 (13)  &49 (43)\\
Mt.\ Maidanak, Uzbekistan &150 &59 (42)&336 (311)&112 (107)&450 (417)&237 (219)&1194 (1096)\\
Abastumani, Georgia (FSU) &70  &0&0&0&41 (32)&0&41 (32)\\
Crimean, Ukraine          &70  &0&0&17 (16)&19 (18)&19 (19)&55 (53)\\
Skinakas, Crete           &130 &0&24 (17)&27 (23)&28 (27)&26 (25)&105 (92)\\
Tuorla, Finland           &103 &0&0&0&44 (17)&0&44 (17)\\
Vallinfreda, Italy        &50  &0&0&0&1 (1)&0&1 (1)\\
Armenzano, Italy          &40  &0&14 (12)&28 (18)&41 (32)&29 (26)&112 (88)\\
Perugia, Italy            &40  &0&0&2 (0)&4 (1)&4 (4)&10 (5)\\
Michael Adrian, Germany   &120 &0&0&0&7 (5)&0&7 (5)\\
Greve, Italy              &32  &0&0&0&1 (1)&0&1 (1)\\
Torino, Italy             &105 &0&6 (2)&12 (11)&132 (102)&83 (62)&233 (177)\\
Heidelberg, Germany       &70  &0&0&0&4 (2)&1 (0)&5 (2)\\
Calar Alto, Spain         &220 &0&0&0&101 (15)&0&101 (15)\\
Sabadell, Spain           &50  &0&0&0&90 (1)&0& 90 (1)\\
Roque (KVA), La Palma     &35  &0&0&0&180 (95)&0&180 (95)\\
Roque (NOT), La Palma     &256 &12 (7)&14 (13)&14 (13)&23 (23)&17 (15)&80 (71)\\
Roque (TNG), La Palma     &358 &0&0&0&2 (2)&0&2 (2)\\
CASLEO, Argentina         &215 &0&0&55 (55)&59 (59)&0&114 (114)\\
Bell, Kentucky            &60  &0&0&0&10 (10)&0&10 (10)\\
Kitt Peak (SARA), Arizona &90  &0&0&3 (2)&30 (19)&3 (3)&36 (24)\\
Mt.\ Lemmon, Arizona      &100 &0&2 (0)&5 (2)&8 (5)&5 (5)&20 (12)\\
Lowell (Perkins), Arizona &180 &0&1 (1)&3 (2)&4 (4)&3 (3)&11 (10)\\
Mauna Kea, Hawaii         &60  &0&3 (0)&4 (1)&0 (0)&1 (1)&8 (2)\\
\hline
Total                     &    &79 (55)&407 (362)&294 (260)&1287 (896)&442 (395)&2509 (1968)\\
\hline
\end{tabular}
\label{obs_opt}
\end{table*}
\begin{table*}
\centering
\caption{List of participating near-IR observatories by longitude and number
of observations performed in the various bands.}
\begin{tabular}{l c |c c c| c}
\hline
Observatory&$d$ (cm)&$N_J$&$N_H$&$N_K$&$N_{\rm tot}$\\
\hline
Campo Imperatore, Italy   &110 &82 (82)&81 (81)&88 (87)&251 (250)\\
TIRGO, Switzerland        &150 &59 (1) &1  (1) &55  (1)&115 (3)\\
Roque (NOT), La Palma     &256 &6  (6) &5  (5) &8   (8)&19 (19)\\
\hline
Total                     &    &147 (89)&87 (87)&151 (96)&385 (272)\\
\hline
\end{tabular}
\label{obs_ir}
\end{table*}

Data were collected as instrumental magnitudes of the source and
comparison stars in the same field.

The source was calibrated preferentially with respect to Stars 1 2 3 ($UBVRI$
photometry by Smith et al.\ \cite{smi85}).
However, since the source was rather faint, in some cases long exposure times produced
saturation of Star 2 (and Star 3), so that it was necessary to
include other reference stars for the calibration.
Hence, we sometimes also used Stars 6 and C1 calibrated in $VRI$ by Fiorucci et al.\ (\cite{fio98}).
We made several checks to test the variation of the source magnitude value when adopting
different choices of reference stars for the calibration: in general, we saw that the variation is
within a few hundredths of a mag, which is usually inside the errors.

In order to construct meaningful light curves, the observations reported in Table \ref{obs_opt}
were carefully analysed and unreliable data discarded.
This was done by zooming into subsequent very short periods of time and
inspecting all the light curves at the same time. 
The same trend in all bands was required within errors;
bad points were eliminated.
In the case of noisy intranight datasets from the same telescope, some binning over time intervals not
exceeding 20 minutes was performed.
In general, data affected by errors larger than 0.2 mag were not included.

Moreover, we investigated the possibility that our photometric measurements of AO 0235+16 were affected
by the contribution of objects very close to the source.
The complex environment of the source was studied by various authors (Yanny et al.\ \cite{yan89};
Burbidge et al.\ \cite{bur96}; Nilsson et al.\ \cite{nil96}); they noticed that there are a few galaxies
at redshift of 0.524 in the source surroundings, and two of them are very close to AO 0235+16.
One galaxy, probably a normal spiral, is 1.3 arcsec to the east,
but it is very faint and thus does not affect the source photometry,
even when the source is faint.
The object about 2 arcsec to the south is more problematic, since it is brighter than the former one,
and it is also known to be an AGN, possibly interacting with the eastern galaxy.
The photometric contribution of this southern object (hereafter named ELISA: elusive intervening southern AGN)
can affect the source magnitude when it is very faint, especially in the bluer
part of the spectrum. This increasing contribution towards the ultraviolet is mainly due to the fact that the AO 0235+16
spectrum is much steeper than that of ELISA, since AO 0235+16 is strongly absorbed in these bands (see
Sect.\ 4). We thus estimated the magnitude of ELISA from the Mt.\ Maidanak and Hanle frames,
where it is often clearly resolved. There is some dispersion in the data, and it is not possible to recognize any clear
variability trend in the period 2000--2004. The best estimates are:
$U=20.8, B=21.4, V=20.9, R=20.5, I=19.9$.
These values are in fair agreement with $R=20.4 \pm 0.1, R-I=0.7\pm 0.2$ given by Nilsson et al.\ (\cite{nil96}), and with
$m_{450}=21.40 \pm 0.01$ estimated by Burbidge et al.\ (\cite{bur96}).
We then converted the AO 0235+16 magnitudes into fluxes, subtracted the ELISA contribution,
and then converted again into magnitudes.

   \begin{figure*}
   \centering
   \caption{$UBVRI$ light curves of AO 0235+16 during the observing season 2003--2004;
   the XMM-Newton pointing of January 18--19 is marked by the vertical (cyan) line at JD = 2453023.
   The ELISA contribution has been subtracted as explained in the text.}
   \label{ubvri}
   \end{figure*}

In Fig.\ \ref{ubvri}
we present the $UBVRI$ light curves obtained during the WEBT campaign,
after applying both the cleaning and ELISA-subtraction
procedures described above. These are the light curves that will be used for the subsequent analysis.
Data from different telescopes are plotted with different colours and symbols in order
to be able to distinguish the single contributions as well as to 
appreciate how well the datasets agree with one another.

The source was in a rather faint state, but was undergoing noticeable variability,
with a total variation of about 2 mag in the best sampled $R$ band,
and changes of about 1 mag on time scales of about a week.

Average colour indices (and standard deviations) obtained by coupling data acquired by the same telescope
within 20 minutes are:
$B-R=1.72$ (0.12), $V-R=0.76$ (0.10), $R-I=0.99$ (0.09). These averages are obtained from 223, 151 and 296
indices, respectively.

Near-IR observations during the WEBT campaign have been carried out with the
110 cm telescope at Campo Imperatore and the 256 cm Nordic Optical Telescope (NOT).
The source magnitude was calibrated with the Gonz\'alez-P\'erez et al.\ (\cite{gon01}) photometry.
The contribution from ELISA in this frequency regime is negligible.

Figure \ref{rjhk} shows the $JHK$ light curves compared with the $R$-band one.
The near-IR light curves are generally undersampled to allow a detailed statement about a 
correspondence with the optical data, but where sufficient sampling has been achieved, 
they indicate that a close correspondence exists, as expected.
Average colour indices (and standard deviations) obtained by coupling data acquired by the same telescope
within 20 minutes are:
$J-K=2.12$ (0.09), $H-K=1.00$ (0.07).
A rough estimate of optical-NIR colour indices can be obtained by coupling data within 60 minutes:
$R-J=2.50$ (0.23), $R-H=3.60$ (0.29), $R-K=4.58$ (0.34). 

   \begin{figure*}
   \centering
   \caption{$JHK$ light curves of AO 0235+16 during the 2003--2004 observing season
   compared to the ELISA-subtracted $R$-band one (top).
   NIR data from Campo Imperatore are shown by (blue) circles, while (red) triangles
   indicate NIR data from the NOT.}
   \label{rjhk}
   \end{figure*}

$JHK$ data before the campaign were obtained on December 18, 2002 (JD = 2452627) with the 150 cm
Italian national telescope TIRGO, at Gornergrat, Switzerland.
These data yield average values of $J=15.28 \pm 0.06, H=14.36 \pm 0.10, K=13.17 \pm 0.08$.

\subsection{Intraday variability}

On very short time scales, the source displayed clear variability in various occasions.
As an example, well-sampled intranight light curves in $V$ and $R$ bands 
were obtained from November 10 to 14, 2001,
with the 215 cm telescope of the CASLEO, Argentina.
The $R$-band light curve is shown in Fig.\ \ref{casleo}, with the same calibration and
ELISA flux subtraction adopted for the data in Fig.\ \ref{ubvri}.
Clear intranight trends are recognizable on November 10, 12, and 13, 
with smooth variations of about 0.1 mag in a few hours.
On internight time scales, variations as fast as $\sim 0.5$ mag in 1--2 days were registered.

Although this behaviour is not as extreme as observed in other occasions,
such as the $\sim 1.2$ mag brightening in one day observed by Romero et al.\ (\cite{rom00})
in November 1999, or similar fast variations detected in the period 1997--1999
(Raiteri et al.\ \cite{rai01}),
nonetheless it reveals that the source can be very active even during faint states.

   \begin{figure}
   \resizebox{\hsize}{!}{\includegraphics{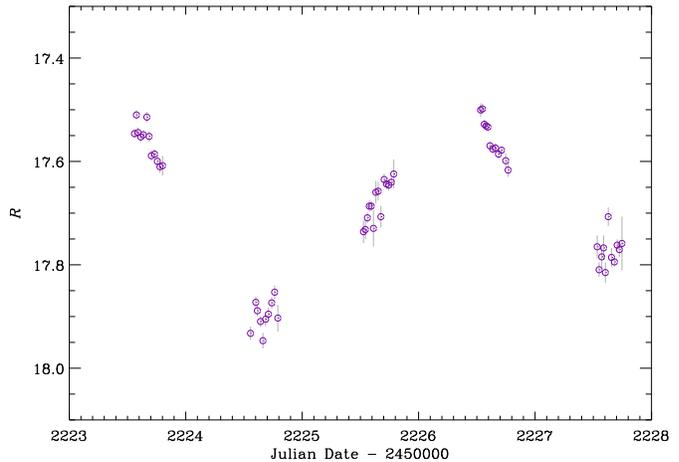}}
   \caption{$R$-band light curve of AO 0235+16 during November 10--14, 2001; data acquired
   with the 215 cm telescope of the CASLEO, Argentina.}
   \label{casleo}
   \end{figure}

   \begin{figure*}
   \centering
   \caption{ELISA-subtracted $R$-band magnitudes of AO 0235+16
   (top) and radio fluxes (Jy) at different frequencies during the WEBT campaign.
   The vertical (cyan) line marks the XMM-Newton pointing of
   January 18--19, 2004. The dotted horizontal line at $F_\nu=1.5 \, \rm Jy$ is drawn to guide
   the eye through the flux variations.}
   \label{radop}
   \end{figure*}

   \begin{figure*}
   \centering
   \caption{ELISA-subtracted $R$-band magnitudes of AO 0235+16 (top)
   compared with radio fluxes (Jy) at different frequencies; in each panel
   the first (red) vertical line marks the last data published by Raiteri et al.\ (\cite{rai01});
   the second (green) line indicates the start of the WEBT campaign, and the third (cyan) one corresponds
   to the XMM-Newton pointing of January 18--19, 2004.}
   \label{radop_long}
   \end{figure*}

\subsection{Radio data}

Long-term radio observations during the WEBT campaign were performed with the 14 m antenna
of the Mets\"ahovi Radio Observatory in Finland, the 26 m paraboloid of the
University of Michigan Radio Astronomy Observatory (UMRAO) in the United States, and the
600 m ring transit radio telescope RATAN-600 of the Russian Academy of Sciences in Russia.
Details on the observing and data reduction procedures can be found e.g.\ in
Ter\"asranta et al.\ (\cite{ter98}), Aller et al.\ (\cite{all99}), and Kovalev et al.\ (\cite{kov99a}).

A comparison among optical magnitudes in the best-sampled $R$ band and radio fluxes at different
frequencies is shown in Fig.\ \ref{radop}.
Data at 37 and 22 GHz are from the Mets\"ahovi Radio Observatory;
data at 14.5, 8.0, and 4.8 GHz are from UMRAO; those at 21.7, 11.1, 7.7, 3.9, and 2.3 GHz are from RATAN.
The two datasets at 37 GHz from Mets\"ahovi have been obtained by two different teams following slightly
different procedures. 
Further data at 4.85 and 10.45 GHz were taken with the 100 m radio telescope at Effelsberg during
the XMM-Newton pointing of January 18--19, 2004 (see Sect.\ 3.2 for more details).
ELISA is not a radio emitting object, so no correction for its contribution is required.

The figure clearly shows a decreasing variability amplitude accompanied by
an increasing base level when going from the higher to the lower radio frequencies.
In general, after an initial small bump and a subsequent lower state,
a quasi-linear trend seems to characterize
the radio light curves starting from JD $\sim$ 2452970, the slope of
which decreases with increasing wavelength, from $4.0 \times 10^{-3}$ at 37 GHz to $-0.3 \times 10^{-3}$ at 5 GHz.
A correlation with the optical data, if it exists, is not obvious.

The analysis of the long-term radio and optical light curves
led Raiteri et al.\ (\cite{rai01}) to the conclusion that the major radio outbursts repeat
quasi-periodically every $\sim 5.7$ years and, when optical data are available,
an optical counterpart is found immediately preceding the radio event.
The next major outburst was predicted to peak in February--March 2004, with a possible window of
$\pm$ 6 months around that date.
As Fig.\ \ref{radop} shows, no major radio or optical outburst was detected during the WEBT campaign.
To make the comparison between the recent and past behaviour of the source possible,
Fig.\ \ref{radop_long} displays the optical and radio light curves over a more extended time period.
The period shown overlaps with that presented in Fig.\ 8 of Raiteri et al.\ (\cite{rai01}),
and includes the last two major radio outbursts which occurred in 1992--1993 and 1998, both of them
presenting an optical counterpart. The first (red) vertical line marks the end of the Raiteri et al.\ 
(\cite{rai01}) data;
the second (green) line indicates the start of the WEBT
campaign on July 1, 2003, and the third (cyan) line corresponds to the XMM-Newton pointing of January 18--19, 2004.
The $R$-band data from Raiteri et al.\ (\cite{rai01}) are here complemented by data from
the Hamburg Quasar Monitoring program (partially published in Schramm et al.\ \cite{sch94})
and by data from CASLEO (Romero et al.\ \cite{rom02});
similarly, past data from RATAN (partially published by Kovalev et al. \cite{kov99b} and
Kiikov et al.\ \cite{kii02}) and data
by Venturi et al.\ (\cite{ven01})
are added to the Raiteri et al.\ (\cite{rai01}) radio light curves.

There are only sparse optical data between the last data presented in the Raiteri et al.\ (\cite{rai01})
paper and the beginning of the WEBT campaign, which show a noticeable mid-term variability, with a
maximum variation of 3.3 mag in 260 days.
Variations of smaller amplitude, but still noticeable, are
found on shorter time scales (see also Sect.\ 2.2). 
These sparse optical data do not allow one to rule out completely
the possibility that a major outburst took place in the 2000--2003.5 period, while the impressive sampling
obtained during the WEBT campaign assures that no major optical outburst occured
in the 2003--2004 observing season.

In contrast, the continuous radio monitoring clearly indicates that there have been no outbursts
since 1998. The new data collected starting from late 2000 show that
the radio flux has been low at all
frequencies, even if not at the minimum levels registered in 1996--1997.

\section{The XMM-Newton pointing of January 18--19, 2004}

The X-ray Multi-Mirror Mission (XMM) - Newton satellite observed AO
0235+16 during revolution number 0753, from January 18, 2004 at 19:06:29 UT
to January 19 at 03:28:24 UT (JD = 2453023.29617--2453023.64472).

\subsection{Results from EPIC}

The European Photon Imaging Camera (EPIC) includes two types of CCD cameras: pn and MOS.
There are two MOS cameras, MOS1 and MOS2, so that EPIC involves 3 detectors in total.

Since an outburst was expected, a medium filter-small window configuration
was chosen for the EPIC detectors in order to avoid both possible contamination
by lower-energy photons and
photon pile-up. Moreover, a small window is also able to reduce the fraction of
out-of-time events recorded by the EPIC-pn detector.

Data were reduced with the Science Analysis System (SAS) software version 6.0.
When extracting the spectrum of both the source and the background, we included the strings
``{\tt FLAG==0}" and ``{\tt PATTERN$<$=4}" in the selection expression for all the three EPIC detectors.
The first string excludes events next to the edges of the CCDs and next to bad pixels
which may have incorrect energies; the second string selects only single and double events, 
which have the best energy calibration.

Source spectra were extracted from a circular region with a 30 arcsec radius\footnote{
We examined an archive image of the AO 0235+16 field taken in August 2000 by Chandra,
which has a space resolution of 0.492 arcsec,
and verified that no X-ray emission was detected from ELISA.
This could be due to intrinsic X-ray faintness, or to Compton absorption, which would favour
a type 2 AGN classification of ELISA.}.
Notwithstanding the small-window configuration, we were able to extract the
background spectrum on the same CCD of the source within circles of 20 arcsec 
radius in the case of MOS1 and MOS2, and 30 arcsec for pn.
These background spectra, when compared with others extracted in different regions,
produce the most satisfactory background subtraction.

The source spectra were grouped with the task ``grppha" of the FTOOL package
in order to have a minimum of 25 counts in each bin.

The detected net count rates are $0.1876 \pm  0.0032$ cts/s for MOS1, $0.1907 \pm 0.0033$
cts/s for MOS2, $0.6261 \pm 0.0063$ cts/s for pn.

In the following we present the results of simultaneous spectral fitting
of the data from the three EPIC detectors performed with the Xspec package version 11.3.0.
Only data in the channels corresponding to energy greater than 0.2 keV were considered.
Once bad channels had been ignored, the channels retained for the subsequent analysis
were: 3--148 for MOS1, 6--153 for MOS2, 7--420 for pn, for a total of 708 PHA bins.

We report the results
in terms of \ion{H}{i} 
column density $N_{\rm H}$ and energy index $\alpha$, which is defined as
$F_E=F_0 E^{-\alpha}$.

The first approach (Model 1) was to fit a simple,
absorbed  power-law model to the data.
The fit was good ($\chi^2/\nu=1.065$, where $\nu$ is the number of degrees of freedom).
The best fit was obtained with $N_{\rm H} = 1.998_{-0.096}^{+0.099} \times 10^{21} \, \rm cm^{-2}$ and
energy index $\alpha=0.692 \pm 0.032$. The unabsorbed 1 keV flux density is
$0.332 \pm  0.011 \, \rm \mu Jy$.

As expected, the value of $N_{\rm H}$ considerably exceeds
the Galactic value of $7.6 \times 10^{20} \, \rm cm^{-2}$ (Elvis et al.\ \cite{elv89}),
indicating extra-absorption.
This extra-absorption had already been found in the previous X-ray spectra, and is ascribed
to a foreground object at $z=0.524$, which causes absorption lines in the optical spectra
(Cohen et al. \cite{coh87}; Nillson et al.\ \cite{nil96}; Raiteri et al.\ \cite{rai05})
\footnote{This absorber is usually identified with
the galaxy 1.3 arcsec to the east,
but a contribution from the ELISA host galaxy, claimed to be interacting with the former,
cannot be ruled out.}.

The next step (Model 2) was to fit the spectra with a power law with Galactic absorption
plus another absorber at $z=0.524$. The fit resulted in a bit lower $\chi^2/\nu$
(1.019), $N_{\rm H} = 2.450_{-0.194}^{+0.201} \times 10^{21} \, \rm cm^{-2}$
at $z=0.524$, and $\alpha=0.645 \pm 0.028$ as best-fit parameters.
The unabsorbed 1 keV flux density in this case is $0.311 \pm  0.008 \, \rm \mu Jy$.

Although both Model 1 and Model 2 are statistically acceptable, we prefer Model 2 since
it provides a plausible explanation for the absorption in excess of the
Galactic value.
Figure \ref{xmm1} shows the data from the three EPIC detectors (top) and the ratio 
between data and folded model (bottom) when applying Model 2.

   \begin{figure}
%   \resizebox{\hsize}{!}{\includegraphics[angle=-90]{2567f6.eps}}
   \caption{Spectra from the three EPIC detectors (top) and ratio between data and folded model (bottom)
   when fitting the data
   with a power-law model with Galactic absorption plus absorption by the foreground galaxy at $z=0.524$.
   The green, black, and red points refer to the pn, MOS1, and MOS2 detectors, respectively.}
   \label{xmm1}
   \end{figure}

Substituting the single power law with a broken power law does not improve the fit.
A more detailed analysis of this spectrum together with other X-ray spectra of this
source will be presented in a forthcoming paper (Raiteri et al.\ \cite{rai05}).

The EPIC light curves do not show any significant variability.

\subsection{Radio-IDV observations with the Effelsberg 100 m telescope}

Densely sampled radio flux-density measurements with the Effelsberg 100 m radio telescope
were performed simultaneously with the XMM-Newton observations of January 18--19,
2004 at $\lambda$2.8 cm and $\lambda$6 cm (10.45 and 4.85 GHz).

On arcsecond scales, AO 0235+16 is a compact radio source and appears
point-like to the Effelsberg 100 m telescope. Its relative brightness of
$\sim 1.7$ Jy at $\lambda$6 cm and $\sim 1.4$ Jy at $\lambda$2.8 cm
in January 2004 allowed the flux-density monitoring to be performed with
``cross scans'' (e.g.\ Heeschen et al.\ \cite{hee87}; Quirrenbach et al.\ \cite{qui92}) 
over the source position. The cross scans consisted of four individual subscans in
azimuthal and elevational back-and-forth direction, respectively, which in
addition were used to control the telescope pointing accuracy throughout the
observations. 
During the measurements, lefthand- and righthand-circular polarization signals
were recorded and later offline combined.
The flux density of AO 0235+16 was measured 7--8 times per
hour at $\lambda$6 cm and once per hour at $\lambda$2.8 cm. Between the
scans on the target source, the two steep-spectrum
secondary calibrators, 3C 67 and 4C 08.10,
were observed, alternately. As primary flux calibrators, the sources 3C 48,
3C 161, and 3C 286 were used (see Ott et al.\ \cite{ott94} and references therein).

Gaussian profiles were fitted to each subscan, yielding the convolution of
point-like source-brightness distribution with the telescope beam.
Small residual pointing
errors were corrected and the amplitudes of the subscans were averaged for
each scan. The secondary calibrators were used to correct for 
time- and elevation-dependent antenna gain effects.
The absolute calibration of the measured flux-density was achieved by scaling
the registered signal from the primary calibrators to match their tabulated
brightnesses. The uncertainties of the flux-density points were estimated from the
residual scatter of the non-variable secondary calibrators, which dominated
over the formal statistical errors. The resulting flux-density uncertainties
are $\sim 0.25$\% at $\lambda$6 cm and $\sim 0.54$\% at $\lambda$2.8 cm.

In Fig.\ \ref{effe_XMM1}, we show the resulting radio light curves.
At both observed frequencies, AO 0235+16 exhibited significant IDV
activity on a level of $\sim 2$--3\%. This corresponds to modulation
indices $m = 0.5$\% at $\lambda$6 cm and $m = 1.2$\%
at $\lambda$2.8 cm, where $m[\%] = 100 \, \sigma_F /$$<$$F$$>$ is defined
via the standard deviation of the flux density, $\sigma_F$, and its average
value in time, $<$$F$$>$.
A higher modulation index at shorter wavelengths, as found here, is not expected if
interstellar scintillation were the physical mechanism
responsible for the IDV activity (e.g.\ Rickett 1986).
This motivates the determination of formal brightness temperatures and Doppler factors from
the assumption that the flux variations have a source-intrinsic origin.
During the observation the flux density of AO 0235+16 decreased by $\sim 30$ mJy
%from $\sim 1.71$ Jy to $\sim 1.68$ Jy,
in $\sim 7$ hours at $\lambda$6 cm, and by $\sim 50$ mJy
%from $\sim 1.45$ Jy to $\sim 1.40$ Jy
in $\sim 6$ hours at $\lambda$2.8 cm.
Both these variations set the same lower limit to the source
brightness temperature\footnote{We adopt
a flat cosmology with $H_0=71 \, \rm km \, s^{-1} \, Mpc^{-1}$ and $\Omega=0.27$,
which yields a source distance of $\sim 6140$ Mpc.}, $T_{\rm b} \ga 10^{17}$ K,
and to the Doppler factor, $\delta \ga 46$,
if the excess over the inverse-Compton
limit of $10^{12}$ K is solely attributed to Doppler boosting.
   \begin{figure}
   \resizebox{\hsize}{!}{\includegraphics{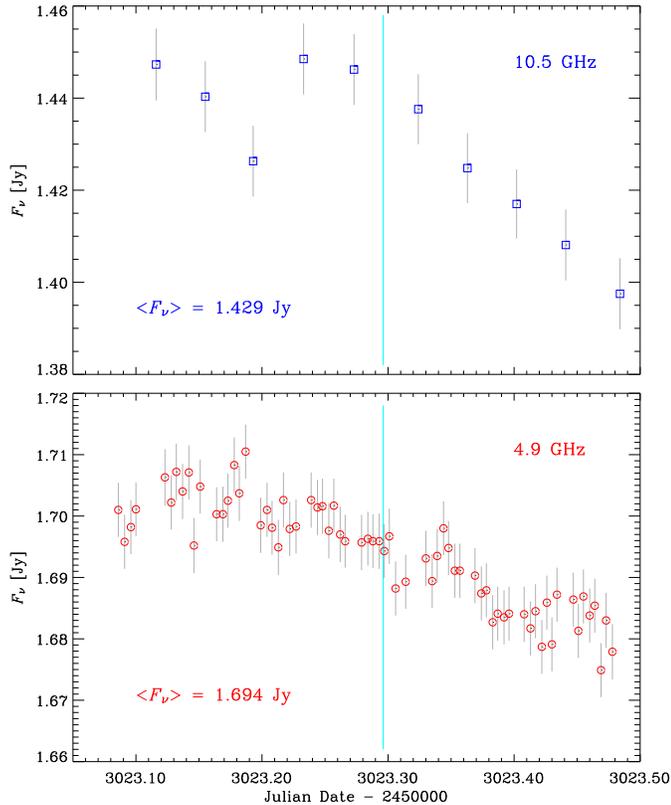}}
   \caption{Radio light curves at $\lambda$2.8 cm (top) and $\lambda$6 cm (bottom)
   obtained with the 100 m radio telescope at Effelsberg
   on January 18, 2004. The vertical line marks the start of the XMM-Newton observations.}
   \label{effe_XMM1}
   \end{figure}

\subsection{Optical observations}

At the time of the XMM-Newton pointing, AO 0235+16 was observable only for few hours
after sunset in the optical band from the typical latitudes of WEBT observers.
Besides this unfavourable situation, weather conditions on January 18--19 were
bad over most of the northern hemisphere, so only a few WEBT collaborators could
observe. Moreover, the source was in one of its faintest states, and this prevented
most of the small telescopes from obtaining utilizable data.

   \begin{figure}
   \resizebox{\hsize}{!}{\includegraphics{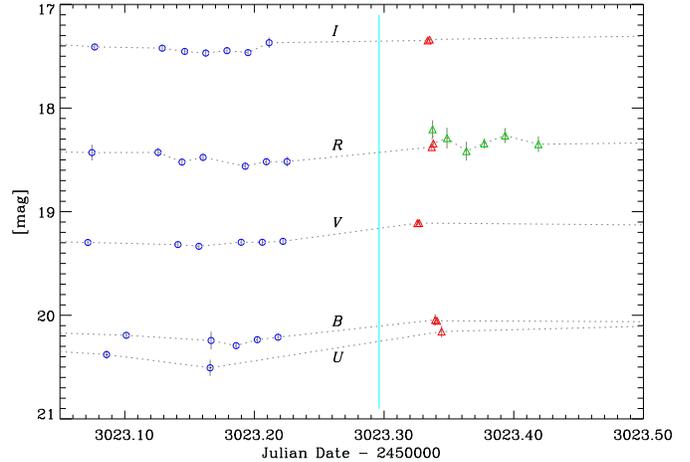}}
   \caption{$UBVRI$ light curves obtained at Mt.\ Maidanak (blue circles) and at La Palma
   by the NOT (red triangles) and by the KVA (green triangles) on January 18, 2004.
   The vertical line marks the start of the XMM-Newton observations.}
   \label{ubvri_xmm1}
   \end{figure}
   
The $UBVRI$ light curves obtained on January 18--19, 2004 are shown in
Fig.\ \ref{ubvri_xmm1}. The observations started with the 150 cm telescope on Mt.\ Maidanak 
and then frames were taken in La Palma simultaneously with the XMM-Newton pointing
by the 256 cm NOT and 35 cm KVA telescopes. 
However, the data uncertainties and poor sampling prevent one to recognize clear variability trends
to be compared with radio and X-ray observations.

\subsection{Results from the Optical Monitor (OM)}

Besides the X-ray detectors, XMM-Newton also has a co-aligned 30 cm
optical--UV telescope, the Optical Monitor (OM). The instrument is equipped with
optical $VBU$ filters plus ultraviolet UV$W1$ ($\lambda_{\rm eff} = 291 \, \rm nm$),
UV$M2$ ($\lambda_{\rm eff} = 231 \, \rm nm$), and UV$W2$ ($\lambda_{\rm eff} =  212 \, \rm nm$) ones.

The OM observation log during the XMM-Newton pointing of AO 0235+16 on January 18--19, 2004
is shown in Table \ref{OM_log}.

The pipeline products (PP)
give the calibrated magnitudes calculated as $m=m_0-2.5 \log \rm(counts/s)$, where $m_0$
are the zeropoints in the various filters. They are continuously updated as more data are becoming available.
As expected, the UV zeropoints are the most unstable ones.
The zeropoints used for the present observations are reported in Table \ref{OM_res}.

\begin{table}
\centering
\caption{Optical Monitor observation log during the AO 0235+16 pointing.}
\begin{tabular}{l c c c c}
\hline
Filter & Exp.\ (s) & Start            & End \\
\hline
$V$    &    2500   &Jan 18, 19:15:29  &Jan 18, 19:57:09\\
$U$    &    2501   &Jan 18, 20:02:17  &Jan 18, 20:43:58\\
$B$    &    2660   &Jan 18, 20:49:03  &Jan 18, 21:33:23\\
$W1$   &    4999   &Jan 18, 21:38:50  &Jan 18, 23:02:09\\
$M2$   &    4998   &Jan 18, 23:07:18  &Jan 19, 00:30:36\\
$W2$   &    5001   &Jan 19, 00:35:43  &Jan 19, 01:59:04\\
Grism1 &    5000   &Jan 19, 02:04:10  &Jan 19, 03:27:30\\
\hline
\label{OM_log}
\end{tabular}
\end{table}

\begin{table*}
\centering
\caption{Results from the Optical Monitor observations of January 18--19, 2004; $m_0$ are the zeropoints.}
\begin{tabular}{l c c c c c c}
\hline
       &$V$           &$B$           &$U$           &$W1$          &$M2$          &$W2$\\
\hline
%\multicolumn{7}{c}{Zeropoints}\\
$m_0$  &17.963        &19.266        &18.259        &17.297        &15.810        &14.862\\
\hline
%\multicolumn{7}{c}{Reference Stars}\\
Star 1 &13.100 (0.002)&13.689 (0.001)&13.582 (0.002)&14.217 (0.004)&15.838 (0.019)&16.010 (0.037)\\
Star 2 &12.784 (0.002)&13.649 (0.001)&13.944 (0.003)&14.804 (0.005)&17.563 (0.057)&17.395 (0.101)\\
Star 3 &12.984 (0.002)&13.761 (0.001)&13.927 (0.003)&14.776 (0.005)&17.353 (0.049)&17.226 (0.089)\\
Star 6 &14.056 (0.004)&14.776 (0.003)&14.741 (0.005)&15.463 (0.008)&17.487 (0.054)&17.613 (0.122)\\
Star C1&14.848 (0.007)&15.868 (0.006)&16.488 (0.014)&17.419 (0.027)& --            & --            \\
%Star E1&15.135 (0.008)&16.332 (0.008)&17.374 (0.027)&18.401 (0.060)& -            & -            \\
%Star E2&16.876 (0.031)&18.350 (0.041)&19.506 (0.166)& -            & -            & -            \\
\hline
%\multicolumn{7}{c}{AO 0235+16}\\
%PP     &19.426 (0.248)&20.315 (0.201)&19.780 (0.188)&19.722 (0.173)& -            & -            \\
%AGN    &19.334 (0.248)&20.050 (0.201)&19.604 (0.188)&19.729 (0.173)& -            & -            \\
%IRAF (6")  &19.228 (0.137)&20.050 (0.115)&19.746 (0.112)&19.423 (0.080)&18.968 (0.222)&$>$18.821     \\
%IRAF (1")  &19.244 (0.073)&20.064 (0.065)&20.084 (0.101)&19.954 (0.109)&19.678 (0.213)&  --   \\
AO 0235+16  &19.22 (0.07)  &20.09 (0.07)  &20.12 (0.10)  &19.95 (0.11)  &19.68 (0.22)  &  --   \\
\hline
\label{OM_res}
\end{tabular}
\end{table*}
In order to check the reliability of the PP photometry, we report in Table \ref{OM_res} the OM
PP magnitudes obtained for the reference stars in the field of AO 0235+16.
We notice that the PP $UBV$ magnitudes for the bright Stars 1 2 3 are close to the Smith et al.\
(\cite{smi85}) values, while the PP $BV$ photometry for the fainter Stars 6 and C1
gives magnitudes larger than those obtained
through ground-based observations (Fiorucci et al. \cite{fio98};
Gonz\'alez-P\'erez et al.\ \cite{gon01}).

In order to avoid the inclusion of the ELISA contribution, we reduced the OM frames with the
IRAF ``phot" routine, using an aperture radius of 2 pixels, corresponding to 0.953 arcsec. 
The results, obtained by differential photometry with respect to the reference stars and magnitudes listed in
Table \ref{OM_res}, are shown in the last row of the same table;
the $VBU$ magnitudes are in satisfactory agreement with those derived from the NOT frames ($V=19.11 \pm 0.02$,
$B=20.05 \pm 0.03$, $U=20.16 \pm 0.06$).

\section{Broad-band spectral energy distribution}

Fitting the intrinsic spectral energy distributions (SEDs) of a blazar is
the only way to test the capability of a model to interpret its broad-band emission.
Since blazars are variable sources,
SEDs must be constructed with simultaneous data; moreover, in order to convert observed magnitudes
into intrinsic fluxes, extinction must be taken into account.

In the case of AO 0235+16, a correct estimate of extinction is not an easy task.
The NASA Extragalactic Database (NED) provides Galactic extinction coefficients
for optical and near-IR bands. Galactic extinction for other bands can be derived following e.g.\
Cardelli et al.\ (\cite{car89}), who give extinction laws from far-IR to far-UV as a function of
wavelength and of the parameter $R_V=A_V/E(B-V)$, where $A_V$ is the extinction expressed in magnitudes
in the $V$ band, and $E(B-V)$ is the extinction difference between the $B$ and $V$ bands.
The mean value for the diffuse Galactic interstellar medium is $R_V=3.1$.
Column 2 of Table \ref{ext} reports the Galactic extinction values given by NED for the $UBVRIJHK$ bands
plus the values obtained from Cardelli et al.\ (\cite{car89}) for the OM $W1$, $M2$,
and $W2$ filters\footnote{Had we used the more recent prescriptions by
Fitzpatrick (\cite{fit99}) in the ultraviolet,
we would have obtained a bit lower extinction corrections:
0.453, 0.694, and 0.763 mag for the $W1$, $M2$, and $W2$
bands, respectively.}, setting $R_V=3.1$.
Notice that, according to NED, $E(B-V)=0.079$.

   \begin{figure}
   \resizebox{\hsize}{!}{\includegraphics{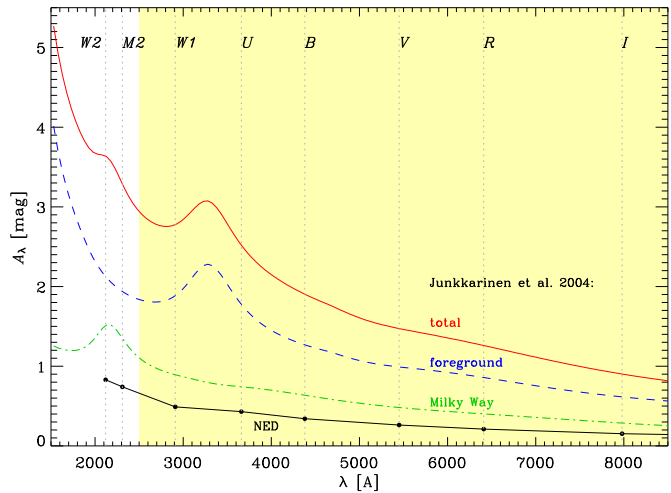}}
   \caption{The amount of absorption (mag) towards AO 0235+16 as a function of
   wavelength. The ``NED" case is compared with the ``Jun" one (see text); in this latter case, the total contribution is
   shown (red, solid line), together with the single contributions by the Milky Way (green, dashed-dotted)
   and by the $z=0.524$ foreground galaxy (blue, dashed). The yellow area indicates the wavelength region where the
   Junkkarinen et al.\ (\cite{jun04}) prescriptions are believed to give a good evaluation of the extinction.}
   \label{junkkarinen}
   \end{figure}

\begin{table}
\centering
\caption{Extinction values (mag) for the Galaxy and the
$z=0.524$ absorber. The ``NED" case adopts the $UBVRIJHK$ values given by the Nasa Extragalactic
Database plus the UV values calculated according to Cardelli et al.\ (\cite{car89}) with $R_V=3.1$.
The ``Jun" case refers to the Junkkarinen et al.\ (\cite{jun04}) prescriptions for the
Galactic as well as the $z=0.524$ absorptions. The last column shows the total extinction
in the ``Jun" case.}
\begin{tabular}{l c c c c}
%\hline
% & NED & \multicolumn{3}{c}{Junkkarinen et al.}\\
\hline
Filter & $A_\lambda^{\rm NED}$ & $A_\lambda^{\rm Jun}$ & $A_\lambda^{\rm Jun}$ &$A_\lambda^{\rm Jun}$\\
       & $z=0$         & $z=0$         & $z=0.524$     & Total\\
\hline
$W2$   & 0.830 & 1.512 & 2.125 & 3.637\\
$M2$   & 0.741 & 1.350 & 1.938 & 3.288\\
$W1$   & 0.490 & 0.893 & 1.884 & 2.777\\
$U$    & 0.430 & 0.742 & 1.777 & 2.519\\
$B$    & 0.341 & 0.636 & 1.268 & 1.904\\
$V$    & 0.262 & 0.482 & 0.991 & 1.473\\
$R$    & 0.211 & 0.402 & 0.858 & 1.260\\
$I$    & 0.153 & 0.287 & 0.615 & 0.902\\
$J$    & 0.071 & 0.140 & 0.318 & 0.458\\
$H$    & 0.046 & 0.088 & 0.187 & 0.275\\
$K$    & 0.029 & 0.055 & 0.116 & 0.171\\
\hline
\label{ext}
\end{tabular}
\end{table}

However, we have already discussed the noticeable extra-absorption affecting the soft part
of the X-ray spectrum, likely due to the $z=0.524$ absorber.
This absorber should affect also the UV--optical--IR spectral region.
The extinction law for this absorber is not directly known, but a possible
recipe can be found in Junkkarinen et al.\ (\cite{jun04}).
These authors performed a number of fits to an HST/STIS spectrum
of AO 0235+16 extending from far-UV to near-IR (1570--10266 \AA).
They assumed that the intrinsic spectrum is a power law (or a broken power law) and
modelled the Galactic extinction according to Cardelli et al.\ (\cite{car89}) with $R_V=3.1$
and $E(B-V)=0.154$. As for the $z=0.524$ absorber, they tried different prescriptions,
including the Cardelli et al.\ (\cite{car89}) laws for the Galaxy with $R_V$ as a free parameter
and the Pei (\cite{pei92}) laws for the Large and Small Magellanic Clouds.
They concluded that the best fit is obtained when considering the Cardelli et al.\ (\cite{car89})
laws with $R_V=2.5$ and $E(B-V)=0.23$.
They warned that the fit is not perfect; deviations occur for $\lambda (z=0) > 8700$ \AA\ and
$\lambda (z=0) < 2500$ \AA; in particular, this recipe leads to an overestimate of the extinction
in the far-UV.
Table \ref{ext} reports the Galactic extinction values following Junkkarinen et al.\ (\cite{jun04}):
when compared with the ``NED" case, the ``Jun" case for $z=0$ (Col.\ 3) implies an almost twice absorption;
moreover, the absorption at $z=0.524$ (Col.\ 4) appears much stronger than the Galactic one.
The total extinction to apply for de-reddening
the source magnitudes in the various bands according to Junkkarinen et al.\ (\cite{jun04})
is given in the last column of Table \ref{ext}.
The same information is plotted in Fig.\ \ref{junkkarinen}: the amount of absorbtion as a function of wavelength
foreseen by the ``NED" case
(black points) is compared with the Junkkarinen et al.\ (\cite{jun04}) prescriptions for the Milky Way (green, dashed-dotted line),
for the $z=0.524$ foreground galaxy (blue, dashed line), and for the total extinction due to both components
(red, solid line). The yellow area highlights the frequency range where these prescriptions lead to a good fit of the HST/STIS
spectrum.

Figure \ref{sed_opt_uv} shows the SED
in the optical--UV frequency range during the XMM-Newton pointing of January 18--19, 2004.
Optical data in $UBVRI$ from the NOT are simultaneous with the optical $VBU$ and ultraviolet
$W1$, $M2$ data from the OM.

De-reddened ground-based optical magnitudes from the NOT were converted into fluxes by using the
zero-mag fluxes by Bessell et al.\ (\cite{bes98}).
These data are shown as (empty and filled) circles in Fig.\ \ref{sed_opt_uv}.
The same was done for the $VBU$ data from the OM; as for the UV bands,
fluxes were obtained using Vega as a reference (see OM
documentation at {\tt http://xmm.vilspa.esa.es}). The OM points are shown as (red) triangles.

In the figure empty symbols represent data obtained by de-reddening for the Galaxy extinction
according to the ``NED" case. Filled symbols indicate the unabsorbed optical--UV SED
when both the Galactic and the $z=0.524$ absorptions are taken into account
following the prescriptions by Junkkarinen et al.\ (\cite{jun04}).

   \begin{figure}
   \resizebox{\hsize}{!}{\includegraphics{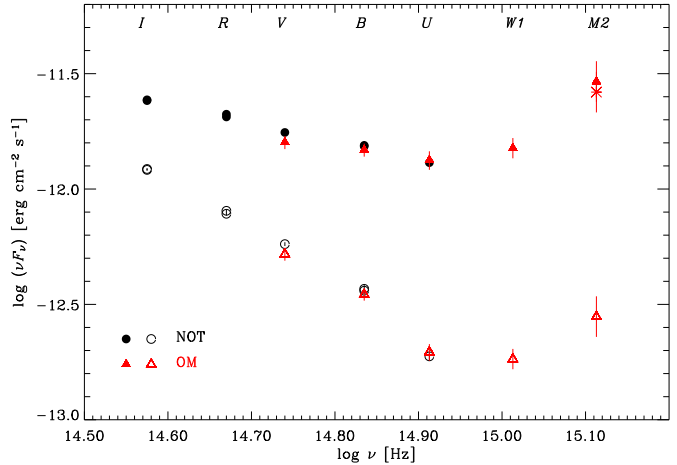}}
   \caption{Spectral energy distribution of AO 0235+16 in the optical--UV frequency range during
   the XMM-Newton pointing of January 18--19, 2004. Empty and filled circles indicate $IRVBU$ data from the NOT
   simultaneous with the XMM-Newton observations; triangles refer to the data from the XMM-Newton Optical Monitor (OM).
   Empty symbols represent data obtained according to the ``NED" case
   for the Galactic extinction; filled symbols refer to the ``Jun" case
   for the Galactic as well as the $z=0.524$ absorptions. 
   The asterisk corresponds to a ``corrected" ``Jun" case (see text).}
   \label{sed_opt_uv}
   \end{figure}

As one can see, when considering also the effect of the $z=0.524$ absorber, the optical spectral
index of AO 0235+16 becomes much smaller (from $\alpha=3.23$ to $\alpha=1.79$ according to the NOT data),
and more similar to that of a typical low-energy peaked BL Lac object.
But the most striking feature is the SED hardening in the ultraviolet, present in both cases.

We warn about use of the $M2$ flux, since it falls in the ``uncertain" region of Fig.\ \ref{junkkarinen},
and, moreover, the calibration of these UV data is also uncertain. Nevertheless, the fact that also in the ``NED" case
we see a dramatic slope change suggests that some slope variation indeed occurs, even if it may be less pronounced.
On the other hand, if we try to quantify the possible overestimate of the absorption in the $M2$ region from the
model of Junkkarinen et al.\ (\cite{jun04}), we find a $\sim 10$\% excess on the flux, 
which would lower the point at the position marked
with an asterisk in Fig.\ \ref{sed_opt_uv}, and would represent only a slight change with respect to the previous situation.

Figure \ref{sed_XMM1} shows the broad-band SED obtained by plotting simultaneous optical--UV--X-ray data
from January 18--19, 2004 and contemporaneous or quasi-contemporaneous radio data.
The optical and UV data are those shown in Fig.\ \ref{sed_opt_uv} (``Jun" case); the X-ray spectrum comes from the Model 2 fit
to the three EPIC detectors together (see Sect.\ 3.1),
taking into account the uncertainties on both the normalization and spectral index.
Radio data at 4.9 and 10.5 GHz are from simultaneous observations at Effelsberg (filled blue diamonds);
upper and lower limits of the variation range during the XMM-Newton pointing have been indicated.
As for the other radio frequencies (empty blue diamonds), the pairs of plotted values correspond to the
data immediately preceding and following the XMM-Newton pointing (see Fig.\ \ref{radop}).
In Fig.\ \ref{sed_XMM1} we have also plotted fictitious $JHK$ data (squares) calculated by adopting the
optical-NIR colour indices obtained from NOT data at a similar brightness level.
They have been de-reddened by using the
extinction values shown in the last column of Table \ref{ext}, 
i.e.\ the same ``Jun" case adopted for the optical--UV bands,
even if this spectral region is outside the HST/STIS spectrum used by Junkkarinen et al.\ (\cite{jun04}) 
to derive their model.

   \begin{figure}
   \resizebox{\hsize}{!}{\includegraphics{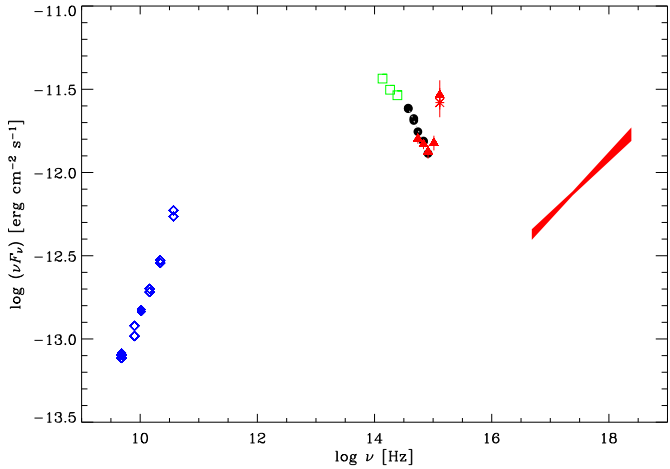}}
   \caption{Broad-band spectral energy distribution of AO 0235+16 during the XMM-Newton pointing of January 18--19, 2004.
   Filled symbols indicate simultaneous data (see text for details).}
   \label{sed_XMM1}
   \end{figure}

\section{Cross-correlations and time delays}

The existence of a correlation between the radio and optical fluxes of AO 0235+16
was discussed in several papers; simultaneous radio and optical
variability was found during the optical outbursts occurred in 1975, 1979, and 1997
(MacLeod et al.\ \cite{mac76}; Ledden et al.\ \cite{led76}; Rieke et al.\
\cite{rie76}; Balonek \& Dent \cite{bal80}; Webb et al.\ \cite{web00}).

Clements et al.\ (\cite{cle95}) analysed time-extended optical and radio light curves,
finding a general delay of the radio events with respect to the optical ones of 0--2 months.
A correlation between the general trends in the two bands was noticed by Takalo et al.\ (\cite{tak98}).
Raiteri et al.\ (\cite{rai01}) compared 25 years of optical and radio data and pointed
out how in some cases optical
outbursts seem to be simultaneous with the radio ones, while in other cases they lead the radio events.
They suggested that this can be the signature of two different variability mechanisms.

Moreover, while Clements et al.\ (\cite{cle95}) did not find any time delay between variations
at different radio frequencies, other authors (O'Dell et al.\ \cite{ode88}; Raiteri et al.\ \cite{rai01})
noticed that the changes observed at the lower frequencies lag the higher-frequency ones.

In the following we re-examine the matter by means of the Discrete Correlation Function
(DCF; Edelson \& Krolik \cite{ede88};
Hufnagel \& Bregman \cite{huf92}; Peterson et al.\ \cite{pet04}), a method
conceived to analyze unevenly sampled data trains. 
For the optical data, we used de-reddened fluxes obtained according to the Junkkarinen et al.\
(\cite{jun04}) prescriptions (see previous section).
Before calculating the DCF we binned the datasets in order
to smooth the intraday features in the optical data and thus obtain a similar sampling
in the radio and optical bands. We tried several bin sizes, from 1 to 10 days, to
check how the results are sensitive to this smoothing procedure.

The DCF run on the new optical and radio data presented in this paper
does not show significant peaks, meaning no radio-optical correlation
over the last four observing seasons, and confirming what
we noticed in Sec.\ 2.3 after a visual inspection of Fig.\ \ref{radop}.

In contrast, a correlation is found when considering the whole datasets,
including the data presented in Raiteri et al.\
(\cite{rai01}), optical data from the Hamburg Quasar Monitoring program 
(partially published by Schramm et al.\ \cite{sch94}) and from CASLEO (Romero et al.\ \cite{rom02}), the
radio data published by Venturi et al.\ (\cite{ven01}), and the old radio data taken with RATAN
(partially published by Kovalev et al.\ \cite{kov99b} and Kiikov et al.\ \cite{kii02}).

In particular, time delays between variations in the best sampled $R$-band light curve and
those in the radio data can now be specified to a higher degree of detail.
Table \ref{dcf} summarizes the results of cross-correlation analysis when the light
curves are binned over 2 days before calculating the DCF;
time lags derived from both the peak,
$\tau_{\rm p}$, and the centroid, $\tau_{\rm c}$, of the DCF are given, where centroids are
obtained by considering all points with ${\rm DCF} > k \, \rm DCF_{p}$, $\rm DCF_{p}$ being the DCF
value of the peak and $k$ ranging from 0.75 to 0.85.
\begin{table}
\centering
\caption{Results of the DCF analysis on the optical ($R$-band) and radio light curves.
Time lags derived from both the peak ($\tau_{\rm p}$) and the centroid ($\tau_{\rm c}$)
are given.}
\begin{tabular}{l c c c}
\hline
Bands & $\tau_{\rm p}$ (days)& DCF$_{\rm p}$ & $\tau_{\rm c}$ (days)\\
\hline
$R$ - 37 GHz   & 30     & 0.4      & 30--35 \\
$R$ - 22 GHz   & 30     & 0.8      & 35--39 \\
$R$ - 14.5 GHz & 30     & 0.6      & 45--50 \\
$R$ - 8 GHz    & 0, 60  & 0.6, 0.7 & $-10$, 60--69\\
$R$ - 5 GHz    & 80     & 0.5      & 65--70\\
\hline
\label{dcf}
\end{tabular}
\end{table}
From the $\tau_{\rm c}$ values shown in the table, one can see that the radio lag
%30--35 days at 37 GHz,
%35--39 days at 21.7--22 GHz,
%45--50 days at 14.5 GHz,
%60--69 days at 8 GHz,
%and 65--70 days at 5 Ghz.
%Hence, the time delay 
seems to increase with decreasing frequency.
Notice that when considering the best-sampled radio light curve at 8 GHz (see Fig.\ \ref{dcf_r8}),
another important peak appears at zero time lag, possibly suggesting that some of 
the radio and optical variations can indeed be simultaneous, as found in some previous studies.
   \begin{figure}
   \resizebox{\hsize}{!}{\includegraphics{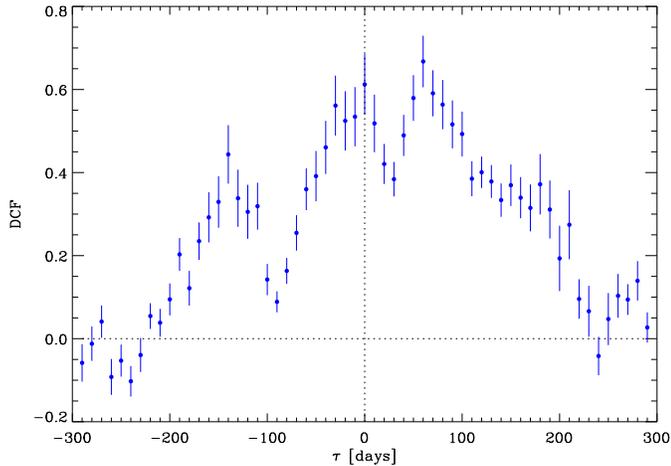}}
   \caption{Discrete correlation function (DFC) between the $R$-band and 8 GHz light curves.
   Two major peaks are visible, corresponding to time lags of 0 and 60 days.}
   \label{dcf_r8}
   \end{figure}

Consistent estimates of delays are also derived from cross-correlating radio datasets among themselves:
taking the 8 GHz light curve as the reference one, its variations seem to follow those seen at 37, 22, 
and 14.5 GHz by 24--35, 29--38, and 14--15 days, respectively, and to lead those at 5 GHz by 10--14 days.
These results are similar to those recently found for other two blazars: S5 0716+71 (Raiteri et al.\ \cite{rai03})
and BL Lac (Villata et al.\ \cite{vil04b}).

The conclusion of the above analysis
is that the radio-optical correlation comes mainly from the major events, which are seen
first in the optical band and then progressively later in the lower-frequency bands, while
during faint states the radio and optical radiations behave in a different way.

\section{Discussion and conclusions}

A huge international observing effort led by the WEBT collaboration
was undertaken to investigate the emission behaviour of the
BL Lac object AO 0235+16 during the observing season 2003--2004.
Observations were carried out at different frequencies, from the radio band to X-ray energies.
Notwithstanding the faintness of the source, which made observations rather difficult,
we obtained an unprecedented monitoring density
in the optical and near-IR bands.

The source was fairly active in the optical and near-IR bands,
but rather ``quiet" at the radio frequencies. 
In particular, the radio flux base level is higher at the lower frequencies.
%Harri: during the quiet stage the spectra of AO 0235+164 is steeper at radio frequencies than during the outburst stage
While the optical and near-IR variations seem to be well correlated,
no correlation is seen between the optical--NIR and radio behaviours.
However, this source has shown correlated optical and radio outbursts in the past,
with short time delays, repeating quasi-periodically every $5.7 \pm 0.5$ years (Raiteri et al.\ \cite{rai01}).
Moreover, in a previous study on the optical and radio behaviour of this source,
Takalo et al.\ (\cite{tak98}) noticed how the radio variability in general follows the optical behaviour,
even if some optical events have no radio counterpart.
Also during the 2003--2004 observing season the main trend seems to be optical activity not significantly
affecting the radio band. 
In other words, the correlation between radio and optical variations found in previous studies
concerns the major outbursts only.

This situation suggests that different variability mechanisms are at work.
As already noticed for other blazars (e.g.\ BL Lac, Villata et al.\ \cite{vil04a}, \cite{vil04b}; S5 0716+71,
Raiteri et al.\ \cite{rai03}), long-term variability can be ascribed to Doppler factor variations (of possible
geometrical origin), while faster flares are most likely of an intrinsic nature.
In the case of AO 0235+16 it has been shown (Ostorero et al.\ \cite{ost04}, see also Romero et al.\ \cite{rom03})
that a rotating helical jet generated by a binary black hole system can account for the periodic behaviour
of radio and correlated optical outbursts. Another scenario suggests that contemporaneous radio and optical outbursts
are the consequences of microlensing due to intervening objects (Stickel et al.\ \cite{sti88}, 
but see Kayser \cite{kay88} for a critical discussion).
On the other hand, the ``inter-outbursts" strong optical activity with no radio counterpart
would imply that the optically emitting region is spatially distinct from the radio emitting one.
In summary, the two emitting regions should be distant enough to provide uncorrelated intrinsic variations,
but close enough to allow a quasi-simultaneous beaming of both of them as the helical jet rotates.

The radio (and optical) outburst predicted to peak in February--March 2004 (with a $\sim 6$ month uncertainty)
has not yet occurred. The WEBT campaign is still ongoing and will cover the entire 2004--2005 observing season.

The XMM-Newton pointing of January 18--19, 2004 found the source in a very faint state.
Nevertheless, the X-ray spectrum is very well determined in both slope and intensity: it is well fitted
by a single power law with extra-absorption. Its hardness suggests an inverse-Compton origin of the observed
X-ray emission.
No significant variability was seen in the X-ray light curve, while a decreasing trend
(about 2--3\% in 6--7 hours) is visible
in the contemporaneous Effelsberg radio data. This variation, if intrinsic, would imply 
a brightness temperature well exceeding the Compton limit, 
and a corresponding Doppler factor $\delta \ga 46$.
As for the simultaneous optical data,
their uncertainty and sampling do not allow one to infer any clear trend.

Our work highlights the role played by the source environment.
On the one hand ELISA (the southern AGN) affects the optical--UV
photometry of AO 0235+16,
especially when it is in a faint state and more in the bluer part of the spectrum.
On the other hand, the eastern galaxy, possibly in conjunction with the ELISA host galaxy,
acts as an absorber of the NIR-to-X-ray radiation.
When the AO 0235+16 broad-band spectral energy distribution is constructed with data
corrected for both these effects, it suggests the existence of a bump in the UV spectral region.
Indeed, when comparing their February 11, 1998
HST/STIS de-reddened spectrum with a contemporaneous ASCA X-ray spectrum,
Junkkarinen et al.\ (\cite{jun04}) noticed that a smooth connection of the optical--UV data with
the X-ray ones in the SED is not possible, but a bump in the UV--soft-X-ray energy range
must be invoked. They considered this possibility unlikely, and suggested that the problem may lie
in an incorrect modelling of the $z=0.524$ absorber extinction law.
A more extended study of this problem through the examination of a number of broad-band SEDs of this source
is under way (Raiteri et al.\ \cite{rai05}).

\begin{acknowledgements}
Based on observations obtained with XMM-Newton, an ESA science mission
with instruments and contributions directly funded by ESA Member States and NASA.
Based on observations with the 100 m telescope of
the MPIfR (Max-Planck-Institut f\"ur Radioastronomie) at Effelsberg.
Based on observations made with the Nordic Optical Telescope, operated
on the island of La Palma jointly by Denmark, Finland, Iceland,
Norway, and Sweden, in the Spanish Observatorio del Roque de los
Muchachos of the Instituto de Astrof\'{\i}sica de Canarias.
Based on observations made with the Italian Telescopio Nazionale Galileo (TNG)
operated on the island of La Palma by the Centro Galileo Galilei of the INAF
(Istituto Nazionale di Astrofisica) at the Spanish Observatorio del Roque de los Muchachos
of the Instituto de Astrof\'{\i}sica de Canarias.
Based on observations taken at TIRGO (Gornergrat, Switzerland).
TIRGO is operated by CAISMI-CNR Arcetri, Firenze, Italy.
The Indian Astronomical Observartory at Hanle  is operated by the
Indian Institute of Astrophysics, Bangalore.
This research has made use of data from the University of Michigan Radio Astronomy Observatory,
which is supported by the National Science Foundation and by funds from the University of Michigan.
\mbox{RATAN--600} observations were partly supported by the
Russian State Program ``Astronomy'' (project~1.2.5.1), the Russian
Ministry of Education and Science, and the Russian Foundation for Basic
Research (projects 01--02--16812 and 02--02--16305).
This work was partly supported by the European Community's Human Potential Programme
under contract HPRN-CT-2002-00321, by the Italian Space
Agency (ASI) under contract CNR-ASI 1/R/073/02 and by the Italian MIUR under grant Cofin2001/028773. 
M.B. acknowledges support by NASA through XMM-Newton GO grant NNG 04GF70G.
V.M.\ Larionov and V.A.\ Hagen-Thorn were supported by Russian Federal Program ``Astronomy'',
project 40.022.1.1.1001.
M.K. was supported for this research through a stipend from the 
International Max Planck Research School (IMPRS) for Radio and Infrared
Astronomy at the University of Bonn.
C.M.R. is grateful to E.\ Trussoni, A.\ Capetti, and J.\ Heidt
for useful discussions.

\end{acknowledgements}

\end{document}